%% file: cycles_v3.tex
\documentclass[prb,showpacs,twocolumn,aps,superscriptaddress,a4paper]{revtex4-1}
\usepackage{color}
\usepackage{dcolumn}
\usepackage{amsmath}
\usepackage{graphicx}
\usepackage{latexsym}
\usepackage{amsfonts}
\usepackage{amssymb}
\DeclareGraphicsExtensions{.pdf,.gif,.jpg}

\input{ourmacros}

\begin{document}

\title{Nonadiabatic Van der Pol oscillations in molecular transport}

\author{Alexey Kartsev}
\affiliation{Mathematical Physics, Lund University, 22100  Lund, Sweden}
\author{Claudio Verdozzi}
\affiliation{Mathematical Physics, Lund University, 22100  Lund, Sweden}
\affiliation{European Theoretical Spectroscopy Facility (ETSF)}
\author{Gianluca Stefanucci}
\affiliation{Dipartimento di Fisica, Universit\`{a} di Roma Tor Vergata,
Via della Ricerca Scientifica 1, 00133 Rome, Italy}
\affiliation{INFN, Laboratori Nazionali di Frascati, Via E. Fermi 40, 00044 Frascati, 
Italy}
\affiliation{European Theoretical Spectroscopy Facility (ETSF)}

\begin{abstract}
The force exerted by the electrons on the nuclei of a 
current-carrying molecular junction can be manipulated to engineer 
nanoscale mechanical systems. In the adiabatic regime a peculiarity of these forces 
is negative friction, responsible for Van der Pol oscillations 
of the nuclear coordinates. In this work we study the robustness of 
the Van der Pol oscillations against high-frequency bias and gate voltage. For 
this purpose we go beyond the adiabatic approximation and perform full 
Ehrenfest dynamics simulations. The numerical scheme implements a mixed 
quantum-classical algorithm for open systems and is capable to deal with 
arbitrary time-dependent driving fields. We find that the Van der 
Pol oscillations are extremely stable. The nonadiabatic electron dynamics 
distorts the trajectory in the momentum-coordinate phase space but 
preserves the limit cycles in an average sense. We further show 
that high-frequency fields change both the oscillation amplitudes and 
the average nuclear positions. By switching the fields off at 
different times one obtains cycles of different amplitudes which 
attain the limit cycle only after considerably long times.

\end{abstract}

\pacs{72.10.Bg, 73.63.-b, 63.20.Ry, 63.20.Kr}

\maketitle

\section{Introduction}

Research activity on the interaction between electrons and nuclei 
began more than 
a century ago, and still today continues to stimulate new ideas and to pose  
challenging problems. Some of the open issues in this field go back to
the early studies by Peierls on one-dimensional lattice instabilities,\cite{peierls}
to continue with the works of Feynman, Fr\"olich and Holstein on 
polarons,\cite{mahan}
the study of charge and heat conduction,\cite{ziman} to arrive to present day open
questions about the role of phonons in superconductivity/magnetism for layered
structures,\cite{dagotto} to mention a few significative examples. 
Modern research covers also more fundamental aspects.
The electron-nuclei interaction (ENI) coupling is typically derived  
from the potential energy 
surfaces of the Born-Oppenheimer approximation.  
As the coupling relies on an approximation, there has been a 
significant effort
in constructing a formally exact theory. Progress has been 
made in this context too. The Born-Oppenheimer ansatz 
for the electron-nuclear 
wave-function is exact in both the static~\cite{hunter} and 
time-dependent~\cite{amg.2010} case and hence the  potential energy 
surfaces constitute a very useful concept even in an exact 
treatment.

In the last few decades, a more quantitative approach to the understanding of the ENI
became possible via computer simulations.
For example ab-initio molecular dynamics,\cite{cp1985}
with a mixed quantum-classical time evolution
for electrons and nuclei, was used to study
phenomena as different as lattice
vibrations and melting, vacancy diffusion,
gas-surface dynamics, etc..
Since the advent of nanotechnology,  the
ENI problem has attracted considerable attention in open nanoscale systems out of 
equilibrium as well.\cite{book2,cuevasbook}
Assessing the nature of ENI and its dependence on the device support
in these low-dimensional geometries
is a key ingredient to control the decoherence of carriers,
the effect of thermal dissipation, in other words to
engineer the ENI to increase device efficiency.\cite{b2005}

While the theoretical study of ENI for steady-state quantum transport
has been the subject of large interest,
\cite{emberly,ths2001,vpl2002,bstmo2003,ctd.2004,cng2004,fbjl2004,pfb2005,fpbj.2007,gnr.2007,gnr.2008,hbt.2009,zrsh.2011,awmtk.2012} 
a real-time description of phenomena like, e.g.,
nuclear rearrangement, multi-stability, electro-migration etc., 
have received less attention (examples of work done in this 
less developed area are Refs. [\onlinecite{hbf.2004,hbfts.2004,vsa.2006,SSSBHT.2006,tb.2011}]).
Recently, the discovery of the nonconservative nature of steady-state
forces\cite{dmt.2009,tde.2010} has re-awakened the interest 
in time-dependent phenomena. Two additional types of forces, both linear in the velocity of 
the nuclear coordinates, have been proposed. 
One force stems from the friction induced by particle-hole 
excitations~\cite{hmzb.2010,lhb.2011} and 
the other force is a Lorentz-like force in which the magnetic field is the curl of the 
Berry's vector potential of the Born-Oppenheimer 
approximation.\cite{lbh.2010} All these forces are contained in 
the Ehrenfest dynamics which evolves the electrons 
quantum-mechanically in the classical field generated by the nuclei and, at the same time, 
the nuclear coordinates according to the classical Newton equation 
in which the forces are generated by the nuclei and the 
electrons. Assuming that the nuclear motion is slow on the electronic 
time-scale {\em and} that the electrons are fully relaxed in the 
instantaneous nuclear configuration, one can expand the electronic 
force in powers of the nuclear velocities (and their derivatives). The 
zeroth order term corresponds to the nonconservative steady-state 
force whereas the first-order term corresponds to the sum of the 
friction force and the Lorentz-like force.\cite{bkevo.2011,bkevo.2012} 
We refer to this approximate nuclear dynamics as the Adiabatic 
Ehrenfest Dynamics (AED).
From the explicit expression of the AED forces, either in terms of  
scattering matrices~\cite{bkevo.2011} or nonequilibrium Green's 
functions,\cite{bkevo.2012} one can show that (i) the steady-state 
force is nonconservative only provided that we are at finite bias
and that the number of nuclear degrees of freedom is larger than one, 
(ii) at zero bias the friction force is always opposite to the nuclear velocity
but it can change sign at finite bias (negative 
friction~\cite{bc.2006,hmzb.2010}) and 
(iii) the Lorentz force vanishes if the number of nuclear degrees of 
freedom is one.

In this work we go beyond the AED by evolving both electrons and nuclei 
according to the {\em full} Ehrenfest dynamics (ED). The ED
has so far being employed to study fast vibrational modes in DC 
regimes.\cite{mb.2011,npmc.2011} 
Here, instead, we  break the adiabatic condition in a different way.
We consider the physical situation of heavy nuclei (and hence slow 
vibrational modes) and drive the system out of equilibrium by high 
frequency AC biases or gate voltages. In fact, our scheme can deal 
with arbitrary driving fields at the same computational cost and is 
not limited to the wide band limit approximation for 
the leads. Furthermore, although our scheme can also include several 
vibrational modes, in this first study we consider only one
vibrational mode and focus on one specific 
issue, namely the negative friction force.
The AED predicts the occurrence of limit 
cycles in the nuclear momentum-coordinate phase space. These cycles 
are similar to those of a van der Pol oscillator~\cite{bkevo.2012} 
and imply that a steady-state is not reached. Is this prediction 
confirmed by the full ED? What are the 
qualitative and quantitative differences? How robust are the van der 
Pol oscillations against ultrafast driving fields?  
To anticipate our conclusions, we  confirm 
the existence of limit cycles, even though the shape and, more 
importantly, the period of the oscillations are different from those 
of the AED. Our main finding, however, is that 
these cycles are remarkably stable against  ultrafast driving 
fields for which the electrons are far from being relaxed, and hence 
the AED is not justified. In the next Section we 
discuss the ED and its adiabatic version.
In Section \ref{numsec} we introduce the model Hamiltonian with a 
single vibrational mode and present results on the time-dependent electron 
current, density and nuclear coordinate. Details on the numerical 
implementation can be found in Appendix \ref{numapp}. Our conclusions and 
outlook are drawn in Section \ref{concsec}.

\section{Theoretical framework}
\label{theorysec}

We consider a system consisting of a left (L) and 
right (R) metallic electrode coupled to a central (C) molecular 
junction. The whole system is initially in the ground state and
then driven out of equilibrium by exposing the electrons to an external
time-dependent bias $V_{\a}(t)$ in lead $\a=$L, R and possibly to some 
time-dependent gate voltage $v_{\rm C}(t)$ in C.
We describe the metallic regions L and R by free-electron Hamiltonians 
\be
\hat{H}_{\a}(t)=\sum_{k}(\e_{k\a}+V_{\a}(t))c^{\dag}_{k\a}c_{k\a},
\ee
with $\a={\rm L,R}$. 
In region C the electrons interact with the classical field 
 generated by the nuclear degrees of freedom ${\bf x}=(x_{1},\ldots,x_{N})$
\be
\hat{H}_{\rm C}({\bf x},t)=\sum_{ij=1}^{M}
h_{ij}({\bf x},t)c^{\dag}_{i}c_{j},
\ee
where the sum runs over the $M$ one-electron states of C.
The nuclear Hamiltonian has the general form
\be
H_{\rm cl}(\blp,\blx)=\sum_{\n=1}^{N}\frac{p_{\n}^{2}}{2M_{\n}}+U_{\rm cl}({\bf x}),
\ee
where $\blp=(p_{1},\ldots,p_{N})$ is canonically conjugated to $\blx$ and $U_{\rm cl}({\bf x})$ is 
the classical potential. Finally the metallic electrodes
are connected to C through the non-local 
tunneling operator
\be
\hat{H}_{\rm T}=\sum_{\a={\rm L,R}}
\sum_{ki}\left(T_{k\a,i}c^{\dag}_{k\a}c_{i}+T^{\ast}_{k\a,i}
c^{\dag}_{i}c_{k\a}\right).
\ee
Thus, the full electron Hamiltonian reads
\be
\hat{H}_{\rm el}(\blx,t)=\hat{H}_{\rm C}(\blx,t)+\sum_{\a={\rm 
L,R}}\hat{H}_{\a}(t)
+\hat{H}_{\rm T}.
\ee

\subsection{Ehrenfest dynamics}
We are interested in calculating time-dependent 
density, current and nuclear coordinates. 
In the limit of heavy 
nuclear masses  the nuclear wavefunction is sharply peaked 
around the classical nuclear coordinates. Then,
an expansion around the classical nuclear trajectory 
leads to a Langevin-type (or stochastic) equation.\cite{mb.2011} Ignoring the 
stochastic forces in this equation corresponds to implement the 
ED. Denoting by $|\Q(t)\ket$ the many-electron state at time $t$, 
the ED for electrons and nuclei is governed by the 
equations
\be
i\frac{d}{d t}|\Q(t)\ket=\hat{H}_{\rm el}(\blx(t),t)|\Q(t)\ket,
\label{qd}
\ee
\bea
\frac{d x_{\n}(t)}{d t}&=&
\left[\frac{\de H_{\rm cl}(\blp,\blx)}{\de p_{\n}}+\bra\Q(t)|
\frac{\de \hat{H}_{\rm el}(\blx,t)}{\de p_{\n}}|\Q(t)\ket\right]_{\substack{\blp=\blp(t)\\\blx=\blx(t)}}
\nonumber \\&=&
\frac{p_{\n}(t)}{M_{\n}},
\label{cdx}
\eea
\bea
\frac{d p_{\n}(t)}{d t}&=&-
\left[ \frac{\de H_{\rm cl}(\blp,\blx)}{\de x_{\n}}+\bra\Q(t)|
\frac{\de \hat{H}_{\rm el}(\blx,t)}{\de x_{\n}}|\Q(t)\ket\right]_{\substack{\blp=\blp(t)\\\blx=\blx(t)}}
\nn\\
&=&-\frac{\de U_{\rm cl}({\bf x}(t))}{\de x_{\n}}-\sum_{ij}
\frac{\de h_{ij}(\blx(t),t)}{\de x_{\n}}
\r_{ji}(t),
\label{cdp}
\eea
where in the last equation
\be
\r_{ji}(t)\equiv \bra\Q(t)|c^{\dag}_{i}c_{j}|\Q(t)\ket
\ee
is the time-dependent one-particle density matrix.
Equations (\ref{qd}-\ref{cdp}) are first-order differential equations 
in time. To solve them we need to specify the boundary conditions. As 
the system is initially in equilibrium, $|\Q(0)\ket=|\Q_{g}\ket$ is 
the electronic ground state, $\blx(0)=\blx_{g}$ are the ground-state 
coordinates and ${\bf p}(0)=0$ (we set $t=0$ as the time at which the 
external bias or gate voltage are switched on). The 
coordinates ${\bf x}_{g}$ 
can be calculated from the zero-force 
equation (see right hand side of Eq. (\ref{cdp}))
\be
\frac{\de U_{\rm cl}({\bf x})}{\de x_{\n}}=
-\sum_{ij}
\frac{\de h_{ij}(\blx)}{\de x_{\n}}\bra\Q_{g}|c^{\dag}_{i}c_{j}|\Q_{g}\ket.
\label{fsc}
\ee 
For $t<0$ the Hamiltonian
$\hat{H}_{\rm el}({\bf x},t)$ is a time-independent free-electron Hamiltonian for any 
$\blx$ and hence its ground state $\Q_{g}=\Q_{g}[\blx]$
is the Slater determinant  formed by the occupied one-electron
wavefunctions $\q_{s}=\q_{s}[\blx]$ of energy 
$\e_{s}=\e_{s}[\blx]$. Consequently the ground-state density matrix 
reads
\be
\r_{g,ji}\equiv\bra\Q_{g}|c^{\dag}_{i}c_{j}|\Q_{g}\ket=
\sum_{s}^{\rm occ}\q^{\ast}_{s}(i)\q_{s}(j)
\label{ope}
\ee
where $\q_{s}(i)=\q_{s}[\blx](i)$ is the amplitude of $\q_{s}$ on the $i$-th one-electron 
state of C.
Equations (\ref{fsc},\ref{ope}) constitute a set of 
coupled equation for the unknown ${\bf x}_{g}$ and $\Q_{g}$. 
In Appendix \ref{numapp} we describe a numerical procedure 
to solve these equations for one-dimensional electrodes.

In order to solve the time-dependent and coupled equations 
(\ref{qd}-\ref{cdp}) in practice we extract from 
Eq. (\ref{qd})  an equation for $\r_{ji}(t)$. Since $\hat{H}_{\rm el}$ is a free-electron 
Hamiltonian at all times we have
\be
\r_{ji}(t)=
\sum_{s}^{\rm occ}\q^{\ast}_{s}(i,t)\q_{s}(j,t)
\label{opet}
\ee
where $\q_{s}(i,t)$ is the time-evolved one-electron wavefunction 
which, by definition, fulfills
\be
i\frac{d}{dt}\q_{s}(i,t)=\sum_{j}h_{ij}(\blx(t),t)\q_{s}(j,t)+\sum_{k\a}T_{k\a,i}^{\ast}
\q_{s}(k\a,t)
\ee
with boundary condition $\q_{s}(i,0)=\q_{s}(i)$.
This equation can be further manipulated to express 
the amplitudes $\q_{s}(k\a,t)$ in the electrodes in terms of the 
amplitudes $\q_{s}(i,t'<t)$ in C with times earlier than 
$t$.\cite{ksarg.2005} We then obtain a close set of equations for $\blx(t)$ and 
$\r_{ji}(t)$. This wavefunction approach  
has been proposed in Ref. \onlinecite{vsa.2006}
and has the advantage of not being limited to wide-band leads and/or 
to DC biases. In Appendix \ref{numapp} we provide some numerical details on the time-propagation algorithm.

An alternative, but equivalent, 
method to calculate $\r_{ji}$ is the NonEquilibrium Green's 
Functions (NEGF) technique.\cite{svlbook} The Green's function is 
defined as
\be
G_{ij}(z,z')=\frac{1}{i}\bra\Q_{g}|\callT\left\{e^{-i\int_{\g}d\bar{z}\hat{H}_{\rm el}[\blx(\bar{z}),\bar{z}]}
c_{i}(z)c^{\dag}_{j}(z')\right\}|\Q_{g}\ket
\ee
where $\g$ is the Keldysh contour going from $-\iif$ to $\iif$ and 
back to $-\iif$, and $z$, $z'$ are contour variables. A contour 
variable can either be on the forward branch $(-\iif,\iif)$ or on the 
backward branch $(\iif,-\iif)$ of $\g$. For any real time $t$ we 
denote by $z=t_{-}$ the contour time on the forward branch and by 
$z=t_{+}$ the contour time on the backward branch. The lesser Green's 
function is defined according to
\be
G_{ij}^{<}(t,t')=G_{ij}(t_{-},t'_{+})=i
\sum_{s}^{\rm occ}\q^{\ast}_{s}(i,t)\q_{s}(j,t')
\ee
and hence 
\be
\r_{ji}(t)=-iG_{ij}^{<}(t,t).
\label{rg<}
\ee
For any finite $t,t'$ the lesser Green's function can be written in  
matrix form as 
\be
G^{<}(t,t')=\int_{-\iif}^{\iif}d\bar{t}d\bar{t}'G^{\rm 
R}(t,\bar{t})\S^{<}(\bar{t},\bar{t}')G^{\rm A}(\bar{t}',t')
\label{g<}
\ee
provided that no bound states are present in the spectrum of 
$\hat{H}_{\rm el}$ when $t\ra\iif$.\cite{s.2007} The 
retarded/advanced Green's functions can be calculated from
\bea
\left(i\frac{d}{dt}-h(\blx(t),t)\right)G^{\rm R}(t,t')=\d(t-t')
\nn\\
+\int_{-\iif}^{\iif}d\bar{t}\,\S^{\rm R}(t,\bar{t})G^{\rm R}(\bar{t},t')
\label{gr}
\eea
with boundary condition $G^{\rm R}(t+\eta,t)=-i$, and $G^{\rm 
A}(t,t')=[G^{\rm R}(t',t)]^{\dag}$.  The lesser and retarded components of the embedding 
self-energy appear in Eqs. (\ref{g<},\ref{gr}). These quantities are
completely determined by the parameters in $\hat{H}_{\a}$ and 
$\hat{H}_{\rm T}$ and read
\bea
\S^{\rm R}_{ij}(t,t')&=&-i\th(t-t')\sum_{\a}
e^{-i\f_{\a}(t,t')}
\nn \\
&\times&\sum_{k}T^{\ast}_{k\a,i}T_{k\a,j}e^{-i\e_{k\a}(t-t')}
\label{srt}
\eea
\bea
\S^{<}_{ij}(t,t')&=&i\sum_{\a}e^{-i\f_{\a}(t,t')}
\nn \\
&\times&\sum_{k}f(\e_{k\a}-\m)T^{\ast}_{k\a,i}T_{k\a,j}e^{-i\e_{k\a}(t-t')}
\eea
where $\f_{\a}(t,t')=\int_{t'}^{t}d\bar{t}\,V_{\a}(\bar{t})$, 
$f(\w)= \th(-\w)$ is the zero temperature Fermi function and $\m$ is the 
chemical potential of the system in equilibrium. This set of 
equations provide an alternative way to implement the ED.

\subsection{Adiabatic Ehrenfest dynamics}

Let us now consider the case of slowly varying driving fields. 
As the nuclei are much heavier than the electrons the electronic 
Green's functions $G^{\rm R/A}(t,t')$ and $G^{<}(t,t')$ depend slowly on the 
center-of-mass time $T=(t+t')/2$. In the adiabatic limit 
$G=G_{ss}$ depends only on the time-difference and equals the 
steady-state Green's function of a system with constant bias 
$V_{\a}$, constant gate voltage $v_{\rm C}$ and steady-state 
coordinates $\blx_{ss}$. The steady-state 
coordinates can be determined similarly to the 
equilibrium case. In Eqs. (\ref{fsc},\ref{ope})
we have to replace $\Q_{g}$ by $\Q_{ss}$ where $\Q_{ss}$ is the 
steady-state Slater determinant formed by all 
right-going scattering states with energy below $\m+V_{\rm L}$ and all 
left-going scattering states with energy below $\m+V_{\rm R}$. 
Alternatively we can calculate $\blx_{ss}$ using NEGF. From Eq. 
(\ref{rg<}) the steady-state one-particle density matrix is
\be
\r_{ss,ji}=-i\int\frac{d\w}{2\p}G^{<}_{ss,ij}(\w),
\ee
and from Eq. (\ref{g<})
\be
G^{<}_{ss}(\w)=G^{\rm R}_{ss}(\w)\S^{<}(\w)G^{\rm A}_{ss}(\w).
\label{g<ss}
\ee
At the steady state the solution of Eq. (\ref{gr}) is simply
\be
G^{\rm R}_{ss}(\w)=\frac{1}{\w-h(\blx_{ss})-\S^{\rm R}(\w)}
\label{grss}
\ee
with, see Eq. (\ref{srt}),
\be
\S^{\rm 
R}_{ij}(\w)=\sum_{k\a}\frac{T^{\ast}_{k\a,i}T_{k\a,j}}{\w-\e_{k\a}-V_{\a}+i\eta}.
\ee
Taking into account that the Fourier transform of the lesser 
self-energy is
\be
\S^{<}(\w)=2\p 
i\sum_{k\a}f(\e_{k\a}-\m)T^{\ast}_{k\a,i}T_{k\a,j}\d(\w-\e_{k\a}-V_{\a})
\ee
we can write $\r_{ss}=\r_{ss}(\blx_{ss})$ in terms of $\blx_{ss}$ and 
then determine $\blx_{ss}$ from
the solution of the zero-force equation
\be
\frac{\de U_{\rm cl}(\blx)}{\de x_{\n}}=-
\sum_{ij}\frac{\de h_{ij}(\blx)}{\de x_{\n}}\r_{ss,ji}(\blx).
\ee

For slowly varying fields is therefore convenient to change variables 
and express the Green's functions in terms of $T=(t+t')/2$ and 
$\t=t-t'$. If we Fourier transform the lesser Green's function 
with respect to the relative time
\be
G^{<}(T,\t)=\int\frac{d\w}{2\p}e^{-i\w\t}G^{<}(T,\w)
\ee
then we can rewrite Eq. (\ref{rg<}) as
\be
\r_{ji}(t)=-i\int\frac{d\w}{2\p}G^{<}(t,\w).
\label{rTt}
\ee
To first order in the nuclear velocities Bode et 
al.~\cite{bkevo.2012} have shown that
\bea
G^{<}(t,\w)&=&G^{<}_{ss}+\frac{i}{2}\sum_{\n}\frac{d 
x_{\n}(t)}{d t}\left[
\frac{\de G^{<}_{ss}}{\de\w}\L_{\n}G^{\rm A}_{ss}\right.
\nn\\
&-&\left.G^{\rm R}_{ss}\L_{\n}\frac{\de G^{<}_{ss}}{\de\w}+
\frac{\de G^{\rm R}_{ss}}{\de\w}\L_{\n}G^{<}_{ss}
-G^{\rm <}_{ss}\L_{\n}\frac{\de G^{\rm A}_{ss}}{\de\w}
\right]
\nn\\
\label{g<exp}
\eea
where the matrix $\L_{\n}=\L_{\n}(\blx(t))\equiv\de h(\blx(t))/\de 
x_{\n}$ and all steady-state Green's functions, see Eqs. 
(\ref{g<ss}, \ref{grss}), are calculated in 
$\blx_{ss}=\blx(t)$.
Substitution of Eq. (\ref{g<exp}) into Eq. (\ref{rTt}) and the 
subsequent substitution of $\r$ into Eq. (\ref{cdp}) allows us to 
decouple the electron and nuclear dynamics, since 
\be
\frac{d p_{\n}(t)}{d t}=F_{\rm cl,\n}(t)+F_{ss,\n}(t)+F_{\rm 
fric,\n}(t)+F_{\rm L,\n}(t)
\ee
where 
\be
F_{\rm cl,\n}(t)=-\frac{\de U_{\rm cl}({\bf x}(t))}{\de x_{\n}}
\ee
is the classical force,
\be
F_{ss,\n}(t)=-
\Tr\left[ \L_{\n}(\blx(t))\r_{ss}(\blx(t))\right]
\ee
is the nonconservative steady-state force,
\be
F_{\rm fric,\n}(t)=-\sum_{\m}\g^{(+)}_{\n\m}(\blx(t))\frac{dx_{\m}(t)}{dt}
\ee
is the friction force and 
\be
F_{\rm L,\n}(t)=-\sum_{\m}\g^{(-)}_{\n\m}(\blx(t))\frac{dx_{\m}(t)}{dt}
\ee
is the Lorentz-like force. In the last two equations
$\g_{\n\m}^{(\pm)}=\g_{\n\m}\pm\g_{\m\n}$, with
\be
\g_{\n\m}=\int\frac{d\w}{2\p}\Tr\left[
G^{<}_{ss}\left(\L_{\n}\frac{\de G^{\rm R}_{ss}}{\de\w}\L_{\m}-
\L_{\m}\frac{\de G^{\rm A}_{ss}}{\de\w}\L_{\n}\right)\right].
\ee
All these forces are well defined functions of $\blx$ and therefore 
we can evolve the nuclear coordinates in time without evolving the 
electronic wavefunctions. This is the adiabatic version of the 
ED and relies on the fact that for any $t$ the 
electronic wavefunctions are steady-state wavefunctions (right- and 
left-going scattering states)  of the Hamiltonian 
$\hat{H}_{\rm el}(\blx(t),t)$. The AED is no longer justified if the 
system is perturbed by driving fields varying on a time scale much 
smaller than the nuclear time-scale.

For $V_{\rm L}= V_{\rm R}$ one can show that the curl $\de 
F_{ss,\n}/\de x_{\m}-\de F_{ss,\m}/\de x_{\n} =0$ and hence that the 
steady-state force is conservative. Instead 
for $V_{\rm L}\neq V_{\rm R}$, i.e., when current flows through the molecular 
junction, this property is not guaranteed.\cite{dmt.2009,bkevo.2012} Of course in the presence 
of only one degree of freedom, $\blx=x$, the steady-state force is, 
by definition, conservative and we can define the total potential
\be
U_{\rm tot}(x)=U_{\rm cl}(x)-\int^{x}dx' F_{ss}(x').
\label{utot}
\ee
The minima of this potential corresponds to stable nuclear 
coordinates in the current carrying system. 

We now consider the friction matrix 
$\g_{\n\m}^{(+)}$. 
If we define the spectral function $A(\w)=i(G^{\rm R}_{ss}(\w)-G^{\rm 
A}_{ss}(\w))$ we have
\be
\g_{\n\m}^{(+)}=-i\int\frac{d\w}{2\p}\Tr\left[
G^{<}_{ss}\left(\L_{\n}\frac{\de A}{\de\w}\L_{\m}+
\L_{\m}\frac{\de A}{\de\w}\L_{\n}\right)\right]
\label{friccoeff}
\ee
When $V_{\rm L}= V_{\rm R}=V$ 
the system is in equilibrium at chemical potential 
$\m+V$. Then, from  the fluctuation-dissipation theorem 
$-iG^{<}_{ss}(\w)=f(\w-\m-V)A(\w)$ and hence
\bea
\g_{\n\m}^{(+)}&=&\int\frac{d\w}{2\p}f(\w-\m-V)\frac{\de}{\de\w}
\Tr\left[A\L_{\n}A\L_{\m}\right]
\nn\\
&=&\frac{1}{2\p}\Tr\left[A\L_{\n}A\L_{\m}\right]_{\w=\m+V}.
\eea
The friction matrix is therefore positive definite. This implies that 
with no current the friction force is opposite to the nuclear 
velocities and 
its effect is to damp the nuclear oscillations around a stable 
position. Again this property can be violated in the current carrying 
system, see Refs. \onlinecite{mb.2011,lhb.2011,bkevo.2012} as well as the 
next Section. Finally we observe that the Lorentz-like force vanishes 
for only one nuclear degree of freedom. In the next Section we 
analyze this case and study the interplay between $F_{ss}$ and 
$F_{\rm fric}$ in a current carrying system. This will be done both 
in terms of AED and full ED simulations, to illustrate 
how the adiabatic picture changes under ultrafast driving fields.

\section{Numerical results}
\label{numsec}

\begin{figure}[tbp]
\includegraphics*[width=.48\textwidth]{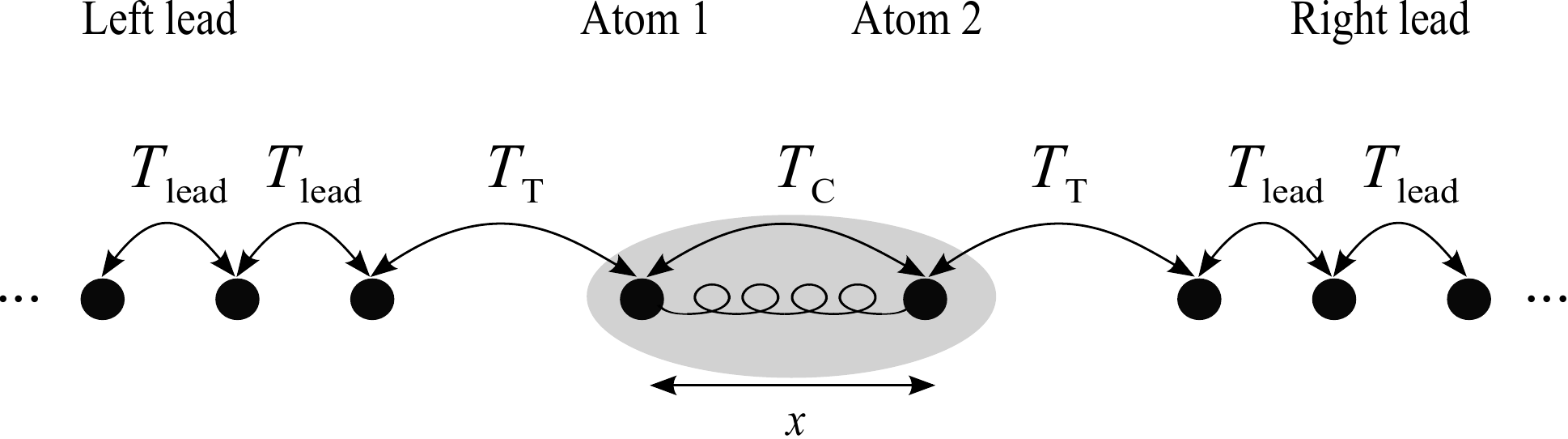}
\caption{Schematic illustration of the two-level molecular junction 
described in the main text.}
\label{system}
\end{figure}

We consider the same model molecular junction as in Ref. 
\onlinecite{mb.2011,bkevo.2012} describing, e.g., 
a polar diatomic molecule and a stretching vibrational mode. 
We assign one single-particle basis function to each 
atom and model the molecule with the $2\times 2$ Hamiltonian
\be
h(x,t)=\left(\begin{array}{cc}
\l x+v_{\rm C}(t) & T_{\rm C} \\
T_{\rm C} & -\l x+v_{\rm C}(t)
\end{array}\right).
\ee
The coordinate $x$ 
moves in the classic harmonic potential
\be
U_{\rm cl}(x)=\frac{1}{2}M\W^{2}x^{2}.
\ee
The junction is coupled through molecule 1 to the left lead and 
through molecule 2 to the right lead, see Fig. \ref{system}. We choose the leads as 
one-dimensional tight-binding metals with nearest neighbor hopping 
$T_{\rm lead}\gg T_{\rm C}$ and zero onsite energy. Thus 
$\e_{k\a}=\e_{k}=2T_{\rm lead}\cos(k)$ with $k\in (0,\p)$. The 
tunneling amplitude from molecule 1 (2) to the left (right) lead is 
denoted by $T_{\rm T}$. 
If we measure all energies in units of $\l^{2}/(M\W^{2})$ then the 
Hamiltonian of region C for electrons and nuclei reads
\be
\hat{H}_{\rm cl}+\hat{H}_{\rm 
C}=\frac{\W^{2}p^{2}}{2\l^{2}}+\frac{\bar{x}^{2}}{2}+T_{\rm 
C}(c^{\dag}_{1}c_{2}+c^{\dag}_{2}c_{1})+\bar{x}(n_{1}-n_{2})
\ee
where $\bar{x}=(M\W^{2}/\l)x$ is a dimensionless coordinate and 
$n_{i}\equiv c^{\dag}_{i}c_{i}$ is
the electron occupation operator on molecule 
$i=1,2$. We consider the following equilibrium parameters: $T_{\rm 
lead}=-10$, $T_{\rm T}=-\sqrt{3}$, $T_{\rm C}=-0.7$
and $\W=0.1$.

\subsection{AED analysis}

We calculate the total potential $U_{\rm tot}$ and 
the friction coefficient $\g^{(+)}$
(in this model the friction matrix is a scalar) in and out of 
equilibrium. For the steady-state values of bias and gate voltage we 
take $v_{\rm C}=0.2$ and $V_{\rm L}=-V_{\rm R}=1$.
Since $T_{\rm lead}\gg T_{\rm C}$  we evaluate the 
embedding self-energy in the Wide Band Limit Approximation (WBLA). 
The WBLA corresponds to taking the limit $ T_{\rm lead},T_{\rm 
T}\ra\iif$ in such a way that $2T_{\rm T}^{2}/T_{\rm lead}=\G$ is a 
finite constant (with our parameter $\G=0.6$).
Then $\S^{\rm R}(\w)=-i\G/2\left(\begin{array}{cc}1 & 0 \\ 0 & 
1\end{array}\right)$ is independent of frequency and 
\be
\S^{<}(\w)=i\G\left(
\begin{array}{cc}
    f(\w-\m-V_{\rm L}) & 0 \\
    0 & f(\w-\m-V_{\rm R})
\end{array}
\right).
\ee

\begin{figure}[tbp]
\includegraphics*[width=.3\textwidth]{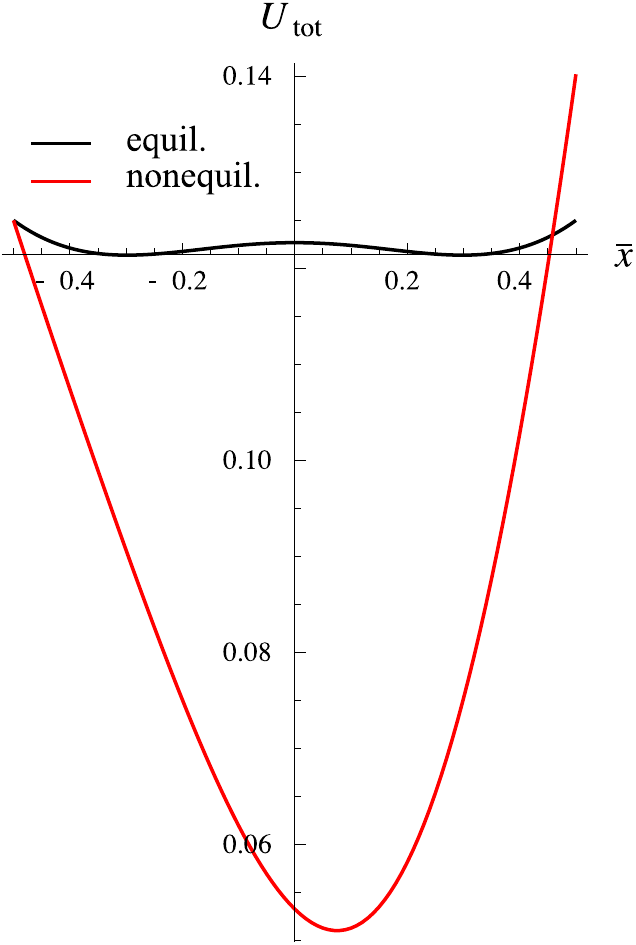}
\caption{(Color online) Total potential as defined in Eq. (\ref{utot}) for the 
equilibrium and nonequilibrium system. }
\label{Utotfig}
\end{figure}

In Fig. \ref{Utotfig} we display the total potential as defined in 
Eq. (\ref{utot}). In equilibrium $U_{\rm tot}(\bar{x})$ exhibits a 
shallow double minimum. The position of the minima corresponds to 
a stable nuclear coordinate. The minima are symmetric around 
$\bar{x}=0$ consistently with the symmetry under reflection of the Hamiltonian. 
In the presence of a bias this reflection symmetry breaks and the current 
carrying system has only one stable coordinate $\bar{x}_{ss}\simeq 
0.088$. The 
nonequilibrium minimum is much deeper than the equilibrium ones 
and occurs at a positive $\bar{x}_{ss}$. From the zero-force equation 
$\bar{x}_{ss}=-(\r_{ss,11}-\r_{ss,22})$, and we infer that the occupation 
on molecule 2 is larger than on molecule 1.

\begin{figure}[tbp]
\includegraphics*[width=.4\textwidth]{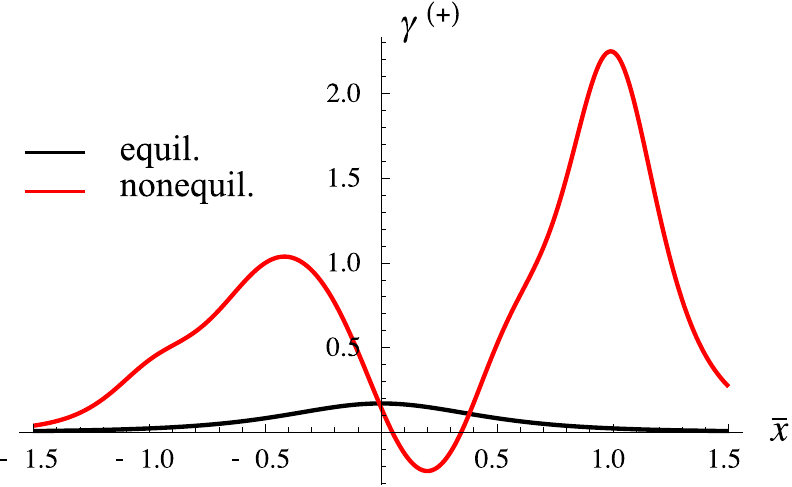}
\caption{(Color online) Friction coefficient as defined in Eq. (\ref{friccoeff}) for the 
equilibrium and nonequilibrium system. $\g^{(+)}$ is in units of 
$M\W$.}
\label{fricfig}
\end{figure}

\begin{figure}[tbp]
\includegraphics*[width=.4\textwidth]{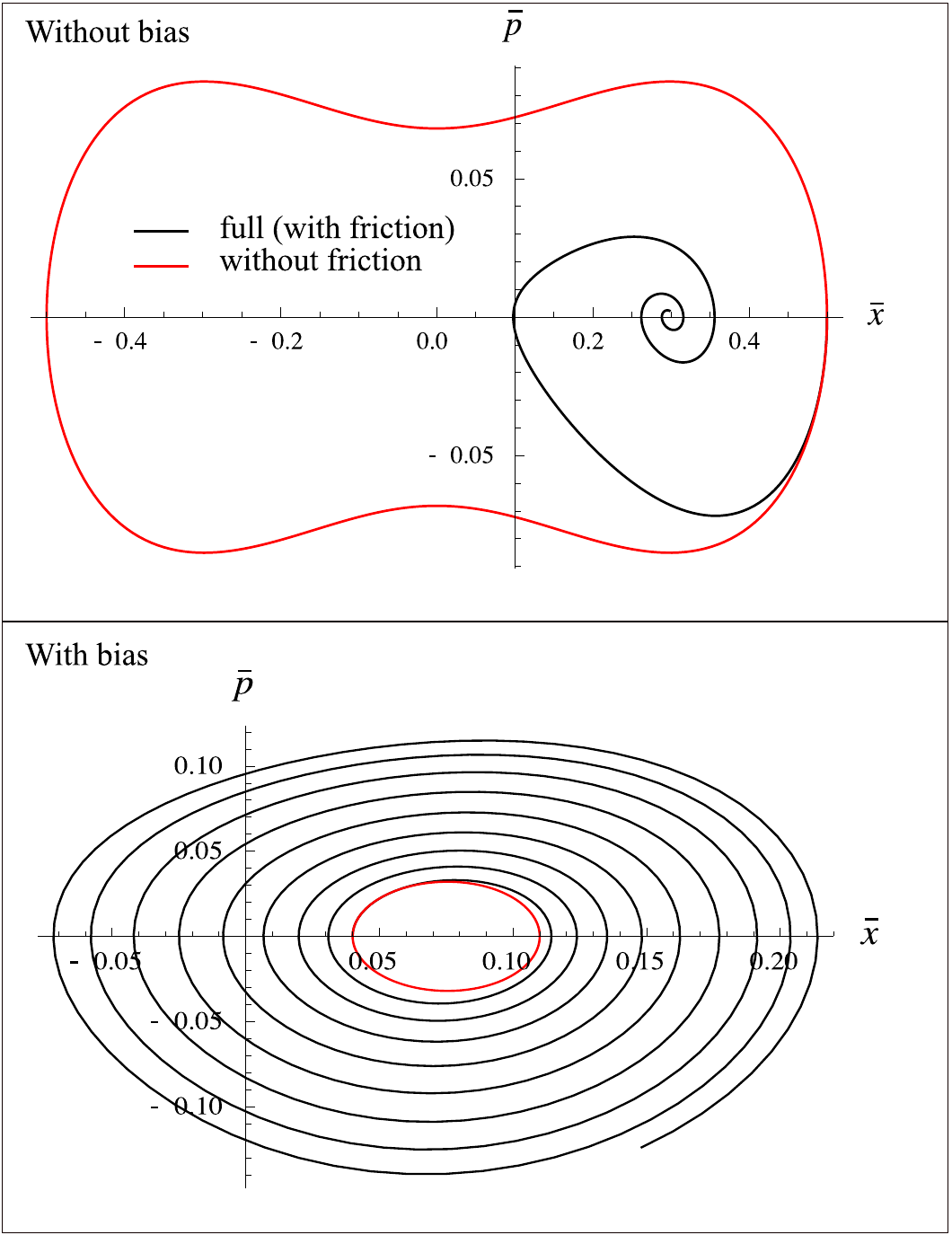}
\caption{(Color online) Solution of Eq. (\ref{aed}) in the $\bar{p}-\bar{x}$ plane. 
Top panel. The system evolves without bias starting from an initial 
nuclear coordinate $\bar{x}(0)=0.5$. Without friction we observe a 
periodic trajectory which explores both minima of the  
potential $U_{\rm tot}$, see Fig. \ref{Utotfig}. In the presence of friction 
(always positive in this case) the nuclear oscillations are damped 
and $\bar{x}$ approaches the positive minimum of $U_{\rm tot}$. 
Bottom panel. The system evolves with a bias $V_{\rm L}=-V_{\rm R}=1$ and 
gate voltage $v_{\rm C}=0.2$ starting from an initial 
nuclear coordinate $\bar{x}(0)=0.04$. Without friction we observe a 
periodic trajectory. Instead with friction the trajectory expand 
outward until reaching a limit cycle.
This is a consequence of the 
negative friction, see Fig. \ref{fricfig}. }
\label{adcycle}
\end{figure}

Next we calculate the friction coefficient $\g^{(+)}$. The results 
are displayed in Fig. \ref{fricfig}. As expected, in equilibrium the friction is 
positive for all values of $\bar{x}$. Instead the nonequilibrium 
friction turns negative in a narrow window of positive $\bar{x}$ 
values. Interestingly, the steady-state coordinate $\bar{x}_{ss}$ 
belongs to this window. This means that if we perform AED simulations 
there is no guarantee 
that a steady-state is reached. For the model that we are considering 
the AED equations reduce to
\be
\frac{d^{2}\bar{x}}{d\bar{t}^{2}}=-\frac{\de U_{\rm 
tot}(\bar{x})}{\de 
\bar{x}}-\g^{(+)}(\bar{x})\frac{d\bar{x}}{d\bar{t}}
\label{aed}
\ee
where $\bar{t}=\W t$. This equation has the same structure as that of 
a van der Pol oscillator $\ddot{y}=-y-\g(y^{2}-1)\dot{y}$ for which 
the function multiplying $\dot{y}$ 
is negative (negative friction) when $y$ is in the range $(-1,1)$ where the 
stable solution $y=0$ lies. As a consequence of this fact one finds 
a limit cycle in the momentum-coordinate phase space. In Fig. \ref{adcycle} 
we show the solution of Eq. (\ref{aed}) in the $\bar{p}-\bar{x}$ 
plane (with $\bar{p}=d\bar{x}/d{\bar t}$)
for a situation in which the 
system has initially a nuclear coordinate $\bar{x}(0)=0.5$ and 
evolves without any bias or gate voltage (top 
panel), and for a situation in which the 
system has initially a nuclear coordinate $\bar{x}(0)=0.04$ 
and evolves in the presence of a bias $V_{\rm L}=-V_{\rm R}=1$ and 
gate voltage $v_{\rm C}=0.2$ (bottom 
panel). For comparison 
we also illustrate the periodic trajectories corresponding to the 
solution of Eq. (\ref{aed}) with $\g^{(+)}=0$. The main difference 
between the two panels is that in the current carrying system the 
nuclear oscillations are not damped. Due to negative friction the 
trajectory expands outward until reaching a limit cycle. We have
checked numerically (not shown) that starting from different 
$\bar{x}$ the trajectory always tend to the same limit cycle.

\subsection{Ehrenfest dynamics simulations}

\begin{figure}[tbp]
\includegraphics*[width=.48\textwidth]{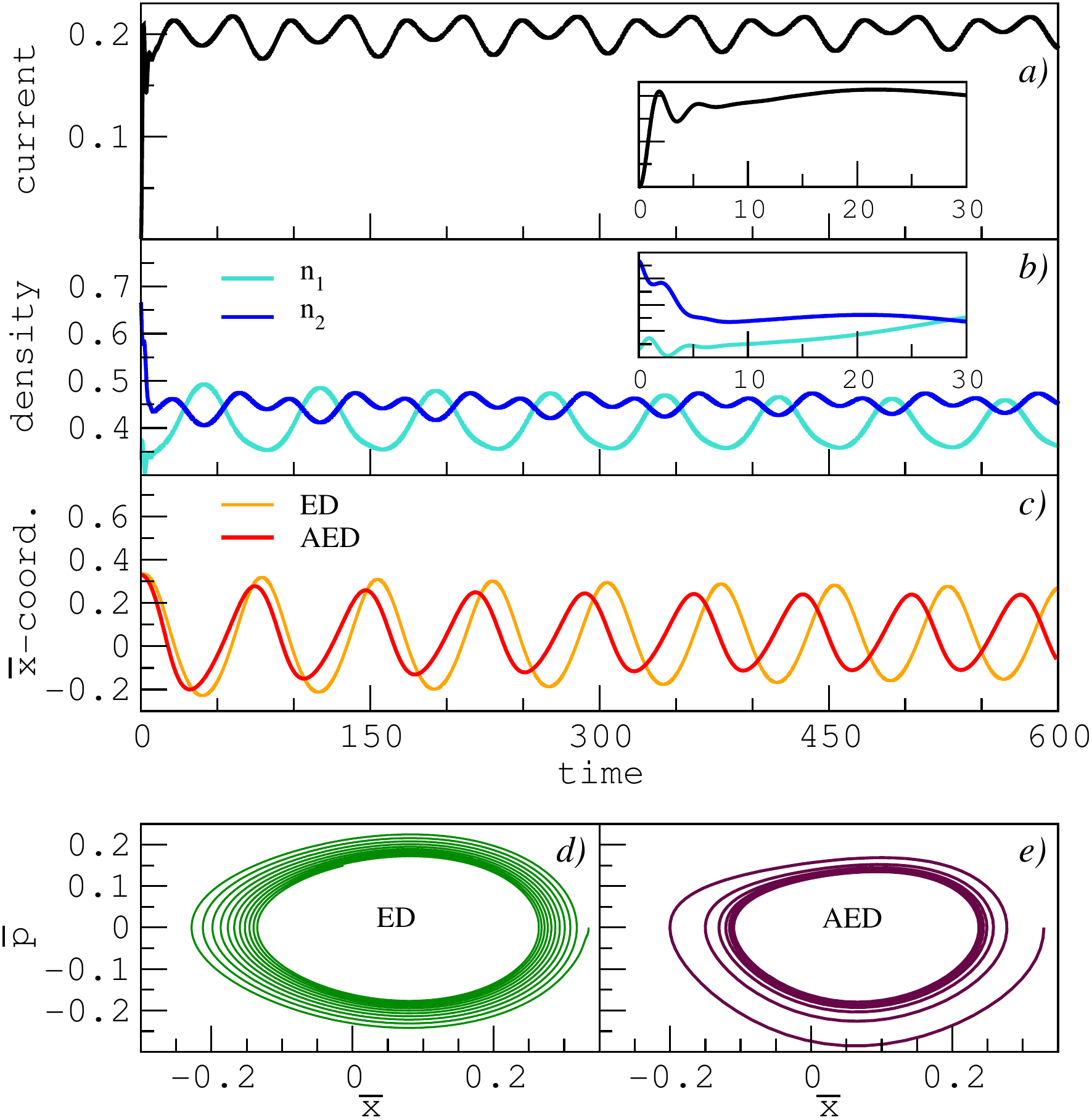}
\caption{(Color online) Results for a sudden switch-on of the external bias and gate 
voltage.
Time-dependent current flowing between atoms 1 and 2 (panel 
a) and occupations of atoms 1 and 2 (panel b). A magnification of the 
initial transient is shown in the insets. Comparison between the 
ED and AED simulations for the nuclear coordinate (panel c).
Trajectories in phase space for the ED (panel d) and AED (panel e) 
simulations. }
\label{sudden}
\end{figure}

We now perform full ED simulations and, instead of 
studying the evolution of the system when the initial coordinate is 
arbitrarily chosen by us, we take the system initially in equilibrium 
and then drive it away from equilibrium using external time-dependent 
biases and/or gate voltages.
In Fig. \ref{sudden} we suddenly switch on  a bias $V_{\rm L}=V_{\rm R}=1$ and 
a gate voltage $v_{\rm C}=0.2$. These are the same parameters as in the 
previous Section. Panels a) and b) show the time-dependent current 
between atoms 1 and 2 and the atomic occupations respectively. After 
a fast transient (see insets) during which the electrons are not 
relaxed,
these quantities start to oscillate on a nuclear time scale. 
Despite the DC bias, no steady-state is reached. 
In panel c) we compare the ED with the AED for 
the $x$ coordinate. In both cases we observe persistent oscillations 
of similar amplitude. However the period of these oscillations is 
different and the curves go out of phase after a few periods.
Also the shape of the oscillations is slightly different. In panels 
d) and e) we put side by side the ED and AED trajectories in phase space. AED 
reaches the limit cycle much faster than the ED.
Apart from these quantitative differences the AED remains a  good 
approximation since during the electronic transient the $x$ coordinate 
moves very little. Thus at times $t\sim 20/\W$ the electrons are 
essentially relaxed in the initial $x$-coordinate. 

\begin{figure}[tbp]
\includegraphics*[width=.48\textwidth]{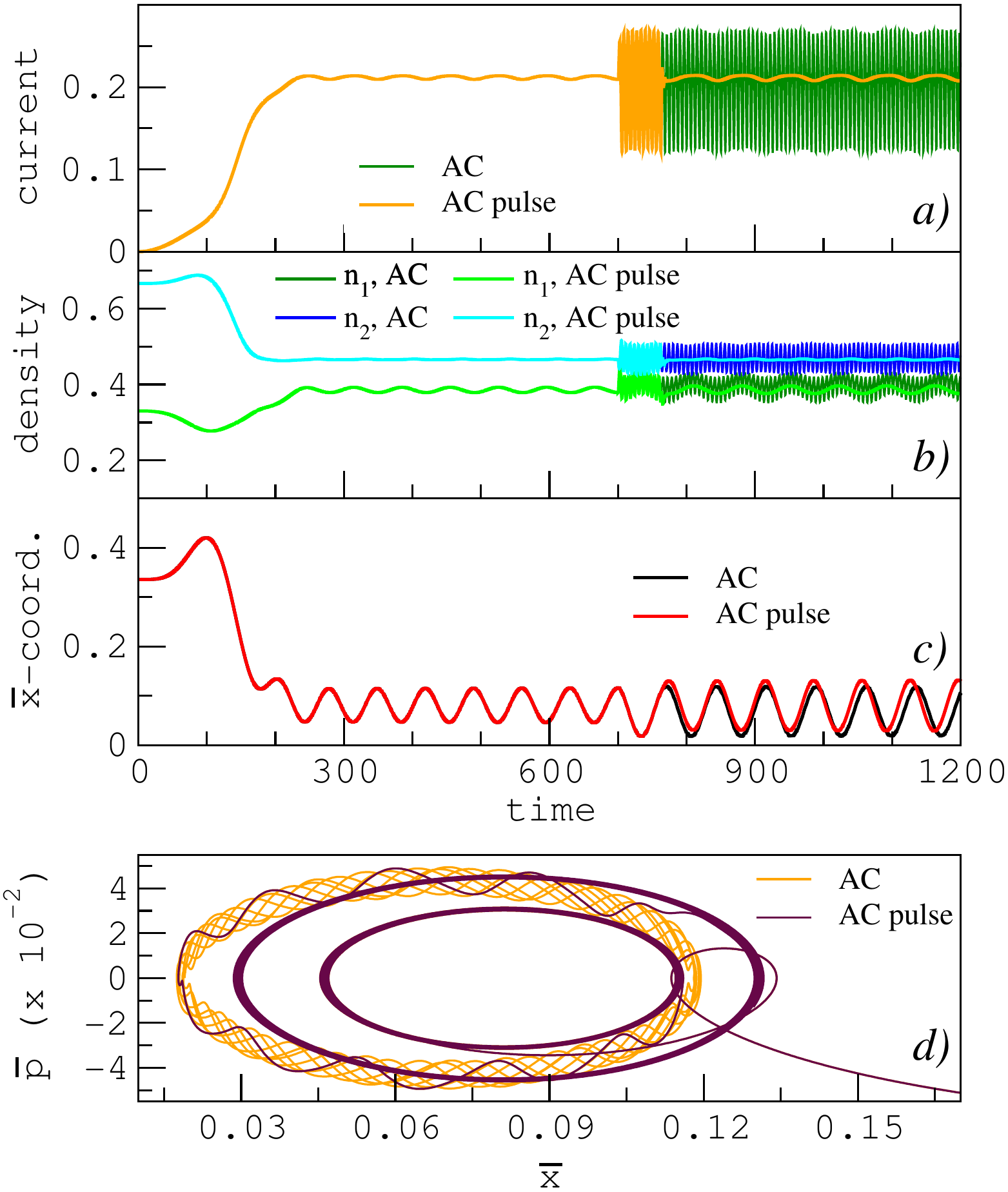}
\caption{(Color online) Results for a smooth switch-on of the DC bias and gate 
voltage followed by the switch-on of a superimposed AC bias.
Time-dependent current flowing between atoms 1 and 2 (panel 
a), occupations of atoms 1 and 2 (panel b) and nuclear coordinate 
(panel c). The curves ``AC pulse'' refer to simulations in 
which the AC bias is switched off after $t=700+2\p\times 10$
(time is in units of $\W^{-1}$).
Trajectories in phase space (panel d).}
\label{acfig}
\end{figure}

New physical 
scenarios may emerge if the electrons are kept away from their 
relaxed state. In this context, a central question is:
Do the  van der Pol oscillations disappear or get distorted? 
To address the issue, we consider two different time-dependent protocols. 
As first protocol,    
we superimpose to the original DC bias a high frequency  AC component.
In order to isolate the effects of the ultrafast AC component, 
we switch on the DC bias smoothly. The explicit form of 
$V_{\rm L}(t)=-V_{\rm R}(t)=V(t)$ is (time is in units of $\W^{-1}$)
\be
V(t)=\left\{
\begin{array}{ll}
    V\sin^{2}(\frac{\p t}{2\times 250}) & \quad t < 250\\
    V & \quad 250<t<700\\
    V+V_{\rm AC}\sin(\w t) & \quad t>700
\end{array}
\right.
\ee
whereas for the gate voltage  $v_{\rm C}(t)=v_{\rm 
C}\sin^{2}(\frac{\p t}{2\times 250})$ for $t<250$ and $v_{\rm 
C}(t)=v_{\rm C}=0.2$ for $t>250$. We maintain the DC component $V=1$ and 
consider 
the amplitude $V_{\rm AC}=0.5$ and the frequency $\w=1$.
The time-dependent current (panel a), 
occupations (panel b)
and nuclear coordinate (panel c) are shown in Fig. \ref{acfig}.
In this figure we also show results (curve ``AC pulse'') of simulations in 
which the superimposed AC bias is switched off after a time 
$t=700+2\p\times 10$. Remarkably the van der Pol oscillations 
persist in this highly nonadiabatic regime. A glance to the 
nuclear coordinate (panel c) would suggest that the AC 
bias is only responsible for increasing the amplitude of the 
oscillations. This is, however, not the case. The trajectory in phase 
space (panel d) reveals that the nuclear coordinate feels the 
nonadiabatic electron dynamics. In fact, we observe cycles with 
superimposed oscillations of the same frequency $\w$ as the AC bias. 
Interestingly an AC pulse can be used to manipulate the radius of 
the cycles. In the ``AC pulse'' curve of panel d) the inner 
cycle sets in 
before the pulse while the outer cycle sets in after the pulse.
The ED is 
nonperturbative in the velocities and their derivatives, and
we are not aware of any mathematical results on the uniqueness of the 
limit cycle. We therefore addressed this issue numerically.
A close inspection to panel d) shows that the inner cycle is moving 
outward whereas the outer cycle is moving inward, thus suggesting the 
uniqueness of the limit cycle even within the ED. We performed several 
simulations with different switching-on protocols of the DC bias and 
 found that cycles with radius larger (smaller) than a critical radius 
move inward (outward). On the basis of this numerical evidence we 
conclude that there exists only one limit cycle within the ED. In 
contrast with the ADE, however, the time to attain the limit cycle is 
considerably longer; hence cycles of different radius can, {\em 
de facto}, be 
considered as quasi-stable limit cycles for practical purposes.

\begin{figure}[tbp]
\includegraphics*[width=.48\textwidth]{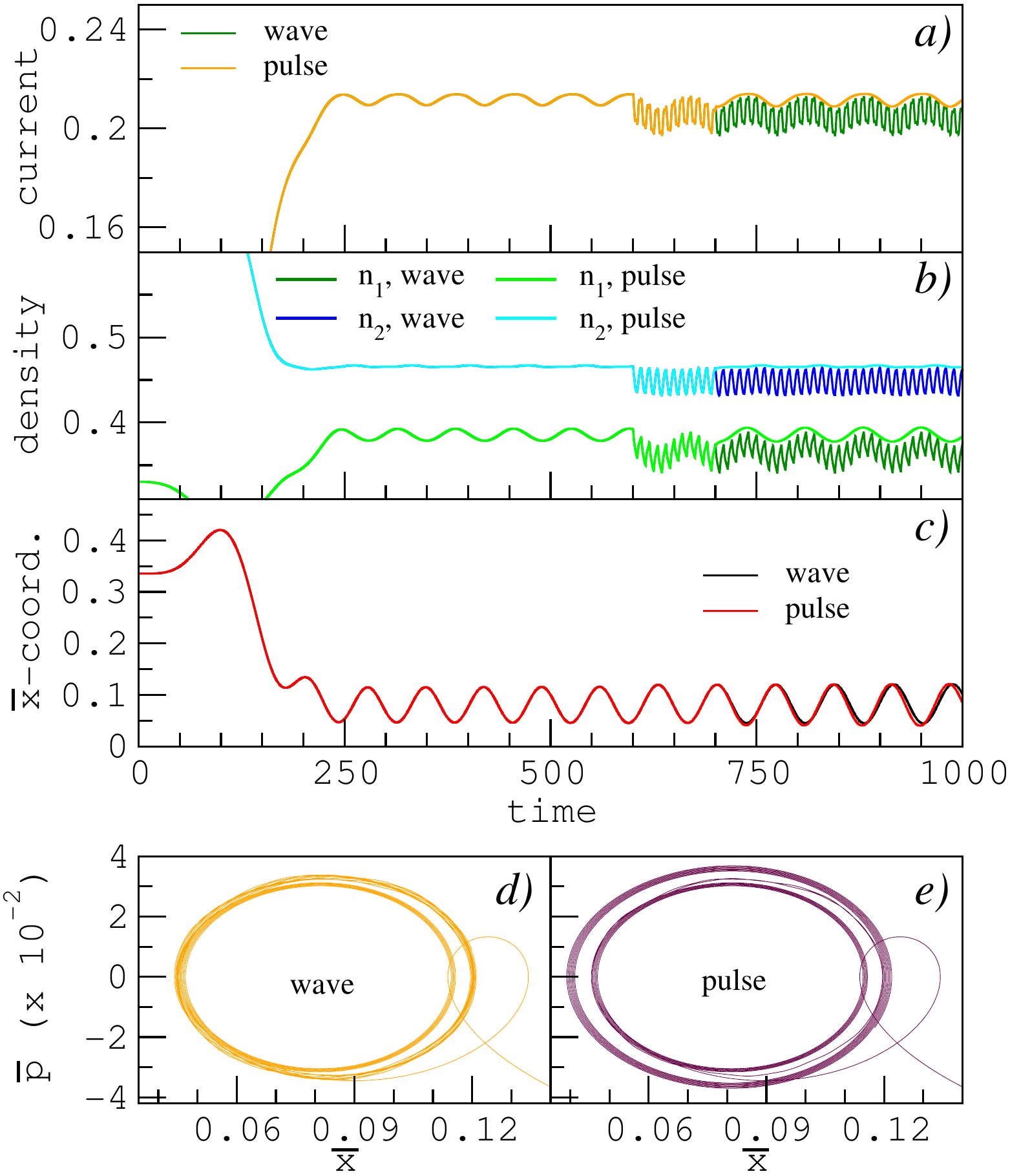}
\caption{(Color online) Results for a smooth switch-on of the DC bias and gate 
voltage followed by the switch-on of a superimposed time-dependent gate.
Time-dependent current flowing between atoms 1 and 2 (panel 
a), occupations of atoms 1 and 2 (panel b) and nuclear coordinate 
(panel c). The curves ``wave'' and ``pulse'' refer to simulations in 
which the time-dependent gate is never switched off and is switched 
off after $t=695$ (time is in units of 
$\W^{-1}$).
Trajectories in phase space for the ``wave'' (panel d) and ``pulse'' 
(panel e) gate.}
\label{gatefig}
\end{figure}

Similar conclusions are reached when the system is perturbed with a second
protocol for a time-dependent perturbation, namely a ultrafast gate voltage. In Fig. \ref{gatefig} we study the response 
of the system to a train of square pulses in the molecular junction. 
After the same 
smooth switching-on of the DC bias and gate as in Fig. \ref{acfig}, we 
superimpose to $v_{\rm C}=0.2$ the time-dependent gate voltage
\be
\d v_{\rm C}(t)=v_{0}\sum_{n=1}^{\iif}S(t-t_{n})
\ee
where $t_{n}=600+n\D$ and $S(t)=1$ if $|t|<\D/4$ and zero otherwise 
(time is in units of $\W^{-1}$). The calculations are performed with 
$\D=10$ and $v_{0}=0.1$.
In the figure the  ``wave'' curves refer to simulations in 
which $\d v_{\rm C}(t)$ is never switched off while the  
``pulse'' curves refer to simulations in which $\d v_{\rm C}(t)$ is switched 
off after a time $t=695$. The van der Pol 
oscillations are stable in both cases. An 
interesting common feature of Figs. \ref{acfig} and \ref{gatefig} is 
that the radius of the quasi-stable limit cycle can be tuned by switching off 
the time-dependent fields at different times. However, if we also 
want to tune the center of the cycle then the time-dependent fields 
have to remain on. Both the ``AC pulse'' curve of Fig. 
\ref{acfig} (panel d) and the ``pulse'' curve of Fig. \ref{gatefig} 
(panel e) exhibit two concentric cycles.
Overall, these features suggest that much more complex  
nuclear trajectories 
 are to be expected when 
the electron dynamics in a junction is nonadiabatic.

\section{Conclusions and outlook }
\label{concsec}

We have studied the robustness of nuclear van der Pol oscillations in 
molecular transport when the junction is subject to ultrafast driving 
fields. In this ultrafast regime the electrons have no time to relax 
and the adiabatic Ehrenfest dynamics (AED) 
is no longer justified. We therefore implemented the full 
Ehrenfest dynamics (ED) using a wavefunction approach. 
The numerical scheme can deal with arbitrary time-dependent 
perturbations at the same computational 
cost and is not limited to wide band leads. We found that the van der 
Pol oscillations are extremely stable. In the DC case the 
AED results are in good qualitative agreement with the full ED 
simulations, as expected. However the ED period of the oscillations 
as well as the damping time to attain the limit cycle are both 
longer than those obtained within the AED. In the presence of 
ultrafast fields the van der Pol oscillations are 
distorted by the nonadiabatic electron dynamics. In all cases we 
observed superimposed oscillations of the same frequency as the 
driving field. We showed that high-frequency biases or gate voltages 
can be used to tune the amplitude of the oscillations and to shift 
the average value of the nuclear coordinate.
By switching the field 
off the amplitude remains large for very long times 
while the average nuclear coordinate 
goes back to its original value rather fast. 
Thus, ultra-fast fields can be used to set in quasi-stable limit 
cycles of desired amplitude. Our numerical evidence suggests that 
every quasi-stable limit cycle eventually attain a unique limit cycle.
We are not aware of any rigorous mathematical proof of this fact. 

In this first work we focused on one aspect of current-induced 
forces, namely the negative friction. In order to observe the 
nonconservative nature of the steady-state force or the Lorentz-like 
force one has to consider at least two vibrational modes. The 
research on current-induced forces is still in its infancy and interesting 
applications like nanomotors have started to appear in the 
literature.\cite{dmt.2009,baj.2008,qz.2009,brvo.2013}
Our results here show that ultrafast fields constitute another knob to tweak  
nanomechanical engines 
and that the theoretical scheme we proposed offers a tool to carry on investigations 
along these lines.

\section*{Acknowledgements}
A.K. and C.V. thank the
EOARD (grant FA8655-08-1-3019) and the ETSF (INFRA-2007-211956) 
for financial support. G.S.  acknowledges funding by MIUR FIRB 
grant No. RBFR12SW0J.

\appendix

\section{Details on the numerical implementation}
\label{numapp}

We consider an arbitrary central region of dimension $M$ described 
by the one-particle matrix $h$. We choose a 
basis set $\{|\f_{i}\ket,\;i=1,\ldots,M\}$ such that an electron can 
hop to the left only through the state $|\f_{1}\ket$ and to the right only 
through the state $|\f_{M}\ket$. (Here and in the following we use 
the Greek letter $\f$ for states strictly localized in region L/R or C and 
$\q$ for states of the entire system S=L+C+R.)
Let $T_{\rm L}$, $T_{\rm R}$ be the corresponding 
hopping parameters (in our model system $T_{\rm L}=T_{\rm R}=T_{\rm T}$). 
The electrodes are described by semi-infinite 
one-dimensional tight-binding models with nearest neighbor hopping 
parameter $T_{\rm lead}=T$ (the same for left and right). The one-particle eigenstates of 
system S can be classified according to their energies. 
The isolated left and right electrodes have a continuous energy spectrum 
between $-2|T|$ and $2|T|$. Therefore, one-particle eigenstates 
with energy $\e<-2|T|$ are bound states with exponential tails in L 
and R. On the other hand, one-particle eigenstates with energy $\e$ 
in the band $(-2|T|,2|T|)$ are extended states delocalized all 
over the system. 

Below we compute the degenerate extended states $\q_{q}^{(a)}$, $a=1,2$, and bound 
states $\q_{b}$ in C. We also describe a damped ground state 
dynamic for the self-consistent solution of Eqs. 
(\ref{fsc},\ref{ope}). Finally we present an efficient algorithm for the 
time-propagation.

\subsection{Extended states}
\label{exst}

Delocalized states are twice degenerate and we  
denote by $\q_{q}^{(1)}$ and $\q_{q}^{(2)}$ the two eigenfunctions 
with eigenenergy $\e_{q}=2T\cos\left(q\right)$, $q\in (0,\pi)$. 
The eigenvalue equation in region C reads
\bea
\sum_{j=1}^{M}\left(\e_{q}\d_{ij}-h_{ij}\right)
\q_{q}^{(a)}(j)=\d_{i,1}T_{\rm L}\q_{q}^{(a)}(\rm L)\;
\nonumber \\ 
+\d_{i,M}T_{\rm R}\q_{q}^{(a)}(\rm R),
\label{sees}
\eea
where $\q_{q}^{(a)}(\a)$, $\a={\rm L,R}$, is the amplitude of the wave function 
on the first site of electrode $\a$ and $\q_{q}^{(a)}(j)\equiv
\bra \f_{j}|\q_{q}^{(a)}\ket$. 
We diagonalize $h$ and find eigenstates
$|\l_{\m}\ket$ with eigenenergies $\e_{\m}$ (the index $\m$ runs 
between 1 and  $M$). In terms of $|\l_{\m}\ket$ and $\e_{\m}$ Eq. (\ref{sees}) 
can be rewritten as
\bea
\q_{q}^{(a)}(i)&=&
T_{\rm L}\q_{q}^{(a)}({\rm L})
\sum_{\m=1}^{M}\frac{\bra\f_{i}|\l_{\m}\ket\bra\l_{\m}|\f_{1}\ket}
{\e_{q}-\e_{\m}}
\nonumber \\ &+&T_{\rm R}\q_{q}^{(a)}({\rm R})
\sum_{\m=1}^{M}\frac{\bra \f_{i}|\l_{\m}\ket\bra\l_{\m}|\f_{M}\ket}
{\e_{q}-\e_{\m}},
\label{escr}
\eea
with $i=1,\ldots,M$. 
This equation allows us to obtain the amplitude of extended states 
in C provided that $\e_{q}\neq \e_{\m}$, $\forall \m$. 
Indeed, we can exploit the degeneracy of $\e_{q}$ and 
choose the vectors 
$\left(\q_{q}^{(1)}({\rm L}),\q_{q}^{(1)}({\rm R})\right)$, 
$\left(\q_{q}^{(2)}({\rm L}),\q_{q}^{(2)}({\rm R})\right)$ as we 
please. Of course, in order to obtain two independent eigenvectors 
$\q_{q}^{(1)}(\a)\neq C \q_{q}^{(2)}(\a)$, $\a={\rm L,R}$, 
with $C$ complex number. 

Having the projection of $|\q_{q}^{(a)}\ket$ onto region C we can match 
it to the analytic form in the leads. We first use the Schr\"odinger 
equation to calculate $\q_{q}^{(a)}(k)$
on sites $k={\rm L}+1,{\rm L}+2$ (second and third sites 
of electrode L) and $k={\rm R}+1,{\rm R}+2$ (second and third sites 
of electrode R). Then we use $\q_{q}^{(a)}(k)$ to compute the 
phase shift and the amplitudes of the oscillation in $\a={\rm L,R}$
\be
\q_{q}^{(a)}(\a+m)=A_{q,\a}^{(a)}\sin\left(qm+\d_{q,\a}^{(a)}\right),
\quad m=1,2.
\ee
The two degenerate eigenfunctions $\q_{q}^{(1)}$ and $\q_{q}^{(2)}$ 
are independent by construction but not orthonormal. 
In order to orthonormalize them we need the overlap 
$N_{aa'}=2\bra\q_{q}^{(a)}|\q_{q}^{(a')}\ket/(\p\d(0))$. It is 
straightforward to show that
\bea
N_{aa'}&=&
A_{q,\rm L}^{(a)}A_{q,\rm L}^{(a')}\cos\left(\d^{(a)}_{q,\rm L}-\d_{q,\rm L}^{(a')}\right)
\nonumber \\ &+&
A_{q,\rm R}^{(a)}A_{q,\rm R}^{(a')}\cos\left(\d^{(a)}_{q,\rm R}-\d_{q,\rm R}^{(a')}\right).
\eea
The wave functions $\q_{q}^{(a)}$ are eventually normalized according to
$\bra\q_{q}^{(a)}|\q_{q'}^{(a')}\ket=2\p\d_{aa'}\d(q-q')$.

\subsection{Bound states} 
\label{bst}

Without loss of generality we choose the hopping parameter $T<0$ in the 
left and right electrodes. Let $|\q_{b}\ket$ be a possible  
bound state of energy $\e_{b}<-2|T|$. As for the extended states, the wavefunction in region C 
is completely determined by the amplitudes $\q_{b}(\a)$, 
$\a={\rm L,R}$ on the first site of electrode $\a$, see Eq. (\ref{sees}).
The amplitudes $\q_{b}(\rm L)$ and $\q_{b}(\rm R)$ 
can be expressed in terms of $\q_{b}(1)=\bra\f_{1}|\q_{b}\ket$ and 
$\q_{b}(M)=\bra\f_{M}|\q_{b}\ket$ 
respectively. We have
\be
\q_{b}({\rm L})=T_{\rm L}\;g(\e_{b})\q_{b}(1),
\quad
\q_{b}({\rm R})=T_{\rm R}\;g(\e_{b})\q_{b}(M),
\label{ir}
\ee
where $g(\w)$ is the retarded Green's function of a semi-infinite chain. 
For $\w<-2|T|$
\be
g(\w)=\frac{\w+\sqrt{\w^{2}-4T^{2}}}{2T^{2}}.
\ee
Using Eqs. (\ref{ir}), the Schr\"odinger equation 
in region C reads
\be
\sum_{j=1}^{M}\left(\e_{b}\d_{ij}-h_{ij}-
\S^{\rm R}_{ij}(\e_{b})\right)
\q_{b}(j)=0,
\label{bsse}
\ee
where the embedding self-energy $\S^{\rm R}(\w)$ is a $M\times M$ matrix 
having only two non-vanishing matrix elements: the (1,1) element which 
is equal to $T_{\rm L}^{2}g(\w)$ and the $(M,M)$ element which is equal to 
$T_{\rm R}^{2}g(\w)$.
Bound-state energies $\e_{b}$ are given by the solutions of 
\be
D(\w)\equiv{\rm Det}\left[
\w-h-\S^{\rm R}(\w)\right]=0, 
\label{det0}
\ee
with $\w<-2|V|$. The corresponding bound state in C can be calculated 
from Eq. (\ref{bsse}). 

In analogy with the procedure described in Section \ref{exst} 
we extended the bound-state wave function
up to the second and third sites of electrode $\a={\rm L,R}$,  
matched it to the analytic form in the leads
\be
\q_{b}(\a+m)=A_{b,\a}e^{-\l_{b,\a}m},\quad
m=1,2,
\label{bsas}
\ee
and calculated the amplitudes $A_{b,\a}$ and penetration lengths
$\l_{b,\a}$. Knowing $|\q_{b}\ket$ in region C and in the leads we 
can calculate 
\be
\bra\q_{b}|\q_{b}\ket=\sum_{j=1}^{M}|\q_{b}(j)|^{2}+
\sum_{\a=\rm L,R}\frac{A_{b,\a}^{2}}{1-e^{-2\l_{b,\a}}}
\ee
and normalize the bound-state wavefunction.

\subsection{Ground state}

The 
parametric dependence of $\hat{H}_{\rm el}$ on the coordinates 
${\bf x}$ renders every eigenstate a function of ${\bf x}$. We 
use the notation $|\q_{q}^{(a)}[{\bf x}]\ket$ and $|\q_{b}[{\bf x}]\ket$ 
for extended and bound states of $\hat{H}_{\rm 
el}[{\bf x}]$. Let us consider the ground state 
$|\Q_{g}\ket=|\Q_{g}[{\bf x}]\ket$ 
of $\hat{H}_{\rm el}[{\bf x}]$. $|\Q_{g}[{\bf x}]\ket$
is a Slater determinant formed by all bound states $|\q_{b}[{\bf x}]\ket$ and 
extended states $|\q_{q}^{(a)}[{\bf x}]\ket$ with energy below the 
chemical potential $\m=2|T|\cos(q_{\rm F})$. The ground 
state value ${\bf x}_{g}$ of the nuclear coordinates can be 
computed from the zero-force equation (\ref{fsc}). 
In our practical implementation
we constructed the one-particle density matrix of Eq. (\ref{ope})
\bea
\r_{g,ji}[\blx]&=&\sum_{b}\q_{b}[\blx]^{\ast}(i)\q_{b}[\blx](j)
\nn\\
&+&
\sum_{a=1}^{2}\int_{0}^{q_{\rm F}}
\frac{d q}{2\p}
\q_{q}^{(a)}[{\bf x}]^{\ast}(i)
\q_{q}^{(a)}[{\bf x}](j)
\eea
and then evolved the coordinates according to the fictitious damped dynamics
\be
M_{n}\ddot{x}_{n}=-\g\dot{x}_{n}-\frac{\de U_{\rm cl}(\blx)}{\de 
x_{n}}-
\sum_{ij}\frac{\de h_{ij}(\blx)}{\de 
x_{n}}\r_{g,ji}(\blx)
\ee
with $\g>0$ some friction coefficient.
 Due to the multivalley nature of 
the potential, the damped dynamics might not converge to the 
lowest-energy solution. We therefore  embedded C in in finite rings 
of increasing length $L$,  
found the energy minimum $\blx_{g}(L)$
and used its extrapolated value 
$\blx_{g}(L\ra\iif)$ as initial condition for the fictitious dynamics.

This concludes the description of the numerical algorithm 
used to find the ground state configuration of 
$\hat{H}_{\rm el}[{\bf x}]+H_{\rm cl}[{\bf p},{\bf x}]$. In the next 
Section we present how to propagate the electronic wavefunctions and 
the nuclear coordinates when 
the system is disturbed by external driving fields.

\subsection{Time evolution}

The evolution is governed by 
Eqs. (\ref{qd},\ref{cdx},\ref{cdp}). 
It is convenient to rearrange the electronic Hamiltonian matrix 
$h_{\rm el}$ of the entire system S=L+C+R as
\be
h_{\rm el}({\bf x}(t),t)=
h_{0}({\bf x}(t),t)+v(t)
\ee
where $h_{0}$ depends on time only through the central region 
\be
h_{0}({\bf x}(t),t)=\left(
\begin{array}{ccc}
h_{\rm L} & h_{\rm LC} & 0 \\
h_{\rm CL} & h({\bf x}(t),t) & h_{\rm CR} \\
0 & h_{\rm RC} & h_{\rm R}^{0}
\end{array}
\right),
\label{hsolo}
\ee
whereas
\be
v(t)=\left[
\begin{array}{ccc}
V_{\rm L}(t) & 0 & 0 \\
0 & 0 & 0 \\
0 & 0 & V_{\rm R}(t) 
\end{array}
\right]
\label{usolo}
\ee
describes the perturbation due to the bias.

For the time-propagation we discretize the time $t_{m}=2m \d$ (where 
$\d$ is an infinitesimal quantity and $m$ is an integer). 
We first propagate all occupied wavefunctions from $t_{m}$ to 
$t_{m+1}$ using the algorithm of Ref. \onlinecite{ksarg.2005}. 
Then we propagate coordinates and momenta from $t_{m}$ to 
$t_{m+2}$ using a Verlet-like algorithm. Finally, we complete the 
propagation of a full time step $\D=4\d$ by evolving the wavefunctions
from $t_{m+1}$ to $t_{m+2}$.
The overall scheme for the time-propagation reads
\begin{widetext}
\begin{equation}
(1+i\d h_{0}^{(m)})\frac{1+i\frac{\d}{2}v^{(m)}}
                            {1-i\frac{\d}{2}v^{(m)}}  |\q_{s}^{(m+1)}\ket=
(1-i\d h_{0}^{(m)})\frac{1-i\frac{\d}{2}v^{(m)}}
                            {1+i\frac{\d}{2}v^{(m)}}  |\q_{s}^{(m)}\ket,
			    \quad\forall\q_{s}\in{\rm occ}
\label{cn1}
\end{equation}
\be
\left\{
\begin{array}{l}
p_{n}^{(m+1)}=p_{n}^{(m)}+2\d F_{n}[{\bf x}^{(m)},\{\q_{s}^{(m+1)}\}]
\\
x_{n}^{(m+2)}=x_{n}^{(m)}+4\d p_{n}^{(m+1)}/M_{n}
\\
p_{n}^{(m+2)}=p_{n}^{(m+1)}+2\d F_{n}[{\bf x}^{(m+2)},\{\q_{s}^{(m+1)}\}]
\end{array}
\right.
\label{v}
\ee
\begin{equation}
(1+i\d h_{0}^{(m+2)})\frac{1+i\frac{\d}{2}v^{(m+1)}}
                              {1-i\frac{\d}{2}v^{(m+1)}}  |\q_{s}^{(m+2)}\ket=
(1-i\d h_{0}^{(m+2)})\frac{1-i\frac{\d}{2}v^{(m+1)}}
                              {1+i\frac{\d}{2}v^{(m+1)}}  |\q_{s}^{(m+1)}\ket,
			      \quad\forall\q_{s}\in{\rm occ}
\label{cn2}
\end{equation}
\end{widetext}
with $m=0,2,4,6,\ldots$. In these equations
$h_{0}^{(m)}=h_{0}(\blx(t_{m}),t_{m})$  and
$v^{(m)}=\frac{1}{2}[v(t_{m+1})+v(t_{m})]$.
We also used the short-hand notation 
$\q_{s}^{(m)}=\q_{s}(t_{m})$, ${\bf x}^{(m)}={\bf x}(t_{m})$ and 
${\bf p}^{(m)}={\bf p}(t_{m})$. The initial values are 
${\bf x}^{(0)}={\bf x}_{g}$, ${\bf p}^{(0)}=0$ 
and $\{\q_{s}^{(0)}\}=\{\q_{s}\}$ ($\{\q_{s}\}$ being the set of occupied 
one-particle eigenstates).
In Eq. (\ref{v}), the force $F_{n}[{\bf x},\{\q_{s}\}]$ depends on the 
coordinates and on the one-particle eigenstates and is 
given by the right hand side of Eq. (\ref{cdp}).

\end{document}

%% file: ourmacros.tex
\newcommand{\be}{\begin{equation}}
\newcommand{\ee}{\end{equation}}
\newcommand{\bea}{\begin{eqnarray}}
\newcommand{\eea}{\end{eqnarray}}

\def\a{\alpha}

\def\g{\gamma}
\def\G{\Gamma}
\def\d{\delta}
\def\D{\Delta}
\def\e{\epsilon}

\def\th{\theta}

\def\l{\lambda}
\def\L{\Lambda}
\def\m{\mu}
\def\n{\nu}

\def\p{\pi}

\def\r{\rho}

\def\S{\Sigma}
\def\t{\tau}
\def\f{\phi}

\def\w{\omega}
\def\W{\Omega}
\def\q{\psi}
\def\Q{\Psi}

\def\blp{{\mathbf p}}

\def\blx{{\mathbf x}}

\def\callT{\mbox{$\mathcal{T}$}}

\def\ra{\rightarrow}

\def\de{\partial}
\def\iif{\infty}
\def\bra{\langle}
\def\ket{\rangle}

\def\Tr{{\rm Tr}}

\def\1op{\hat{\mathbbm{1}}}
\def\nn{\nonumber}